\documentclass[useAMS,usenatbib]{mn2e}
\usepackage{times, graphicx, setspace, subfigure, latexsym, amssymb, amsmath, fancyhdr}
\def\lsim{\mathrel{\rlap{\lower4pt\hbox{\hskip1pt$\sim$}}
    \raise1pt\hbox{$<$}}}                
\def\gsim{\mathrel{\rlap{\lower4pt\hbox{\hskip1pt$\sim$}}
    \raise1pt\hbox{$>$}}}                

\title[HTRU VII: discovery of 5 MSPs]{The High Time Resolution Universe
  Pulsar Survey -- VII:\\discovery of five millisecond pulsars and the different
  luminosity properties of binary and isolated recycled pulsars.}
\author
[M. Burgay et al.]{M. Burgay$^{1}$\thanks{E-mail:burgay@oa-cagliari.inaf.it}, 
M. Bailes$^{2,3}$, S. D. Bates$^{4}$, N. D. R. Bhat$^{2,5}$, S. Burke-Spolaor$^{6}$, 
\newauthor 
D. J. Champion$^{7}$, P. Coster$^{2,8}$, N. D'Amico$^{1,9}$,
  S. Johnston$^{8}$, M. J. Keith$^{8}$, M. Kramer$^{7,10}$, 
\newauthor 
L. Levin$^{4}$, A. G. Lyne$^{10}$, S. Milia$^{1}$, C. Ng$^{7}$, A. 
Possenti$^{1}$, B. W. Stappers$^{10}$, D. Thornton$^{10,8}$, 
\newauthor 
C. Tiburzi$^{1,9}$, W. van Straten$^{2}$, C. G. Bassa$^{10}$
\\
$^{1}$INAF/Osservatorio Astronomico di Cagliari, localit\`a Poggio dei Pini, strada 54, 09012 Capoterra, Italy\\
$^{2}$Centre for Astrophysics and Supercomputing, Swinburne University of Technology, Mail H39, PO Box 218, VIC 3122, Australia\\
$^{3}$ARC Centre of Excellence for All-sky Astrophysics, 44 Rosehill Street Redfern, NSW 2016, Australia \\
$^{4}$Department of Physics, West Virginia University, 111 Hodges Hall, PO Box 6315, Morgantown, WV 26506, USA\\
$^{5}$International Centre for Radio Astronomy Research, Curtin University, Bentley, WA 6102, Australia\\
$^{6}$Jet Propulsion Laboratory, California Institute of Technology, 4800 Oak Grove Drive, Pasadena CA 91104 USA\\
$^{7}$Max Planck Institut f\"{u}r Radioastronomie, Auf dem H\"{u}gel 69, 53121 Bonn, Germany\\
$^{8}$CSIRO Astronomy and Space Science, Australia Telescope National Facility, PO Box 76, Epping, NSW 1710, Australia\\
$^{9}$Dipartimento di Fisica, Universit\`a degli Studi di Cagliari, SP Monserrato-Sestu km 0.7, I-09042 Monserrato, Italy\\
$^{10}$Jodrell Bank Centre for Astrophysics, University of Manchester, Alan Turing Building, Oxford Road, Manchester M13 9PL, United Kingdom\\
}

\begin{document}

\date{}

\pagerange{\pageref{firstpage}--\pageref{lastpage}} \pubyear{2012}

\maketitle
\label{firstpage}

\begin{abstract}
This paper presents the discovery and timing parameters for five
millisecond pulsars (MSPs), four in binary systems with probable white
dwarf companions and one isolated, found in ongoing processing of the
High Time Resolution Universe Pulsar Survey (HTRU). We also present
high quality polarimetric data on four of them. These further
discoveries confirm the high potential of our survey in finding
pulsars with very short spin periods. At least two of these five MSPs
are excellent candidates to be included in the Pulsar Timing Array
projects. Thanks to the wealth of MSP discoveries in the HTRU survey,
we revisit the question of whether the luminosity distributions of
isolated and binary MSPs are different. Using the Cordes and Lazio
distance model and our new and catalogue flux density measurements, we
find that 41 of the 42 most luminous MSPs in the Galactic disk are in
binaries and a statistical analysis suggests that the luminosity
functions differ with 99.9\% significance. We conclude that the
formation process that leads to solitary MSPs affects their
luminosities, despite their period and period derivatives being
similar to those of pulsars in binary systems.

\end{abstract}

\begin{keywords}
stars: pulsars: individual: PSR\,J1431--5740, PSR\,J1545--4550,
PSR\,J1825--0319, PSR\,J1832--0836, PSR\,J2236--5527
\end{keywords}

\section{Introduction}
According to the {\it{recycling}} model (e.g. \citealt{acrs82})
millisecond - or recycled - pulsars (MSPs) are formed in binary
systems where the companion star transfers mass, and hence angular
momentum, onto the neutron star (NS). Extended mass transfer from low
mass companions can cause the NS to be spun up to rotational periods
of at least 1.5 milliseconds. Higher mass systems, transferring mass
over shorter time scales and mostly via a wind \citep{bv91}, cannot
push the recycling process down to periods below $\sim$ 20
milliseconds. At the same time, due to still unclear mechanisms,
possibly linked to the accreted mass itself, the magnetic field B is
decreased down to $10^{8-10}$ G.

About 75\% of all known MSPs in the Galactic field (i.e. outside
Globular Clusters, where exchange interactions can change the final
outcome of the evolution of an MSP) are found in binary systems (data
from the ATNF pulsar
catalogue\footnote{\ttfamily{http://www.atnf.csiro.au/people/pulsar/psrcat/}};
\citealt{mhth05}), supporting the above scenario. The formation path
of isolated MSPs, though, is still a matter of debate: those with spin
periods of few tens of ms or more could derive from double neutron
star systems disrupted during the second supernova explosion
(Disrupted Recycled Pulsars, DRPs; \citealt{cnt96, lma+04}) while the
shorter period pulsars could be the result of a binary system where
the companion was so light that the interaction with the pulsar wind
was able to completely destroy it \citep{bv91}. There are now a few
examples of the so called {\it{black widow}} (BW) pulsar systems
\citep{krst88, pebk88, vv88}, in which the companion star is very
light and in which the radio signal is eclipsed by matter ablated from
the companion for significant parts of the orbit. These objects have
been suggested as possible progenitors of the shorter period isolated
MSPs, although the ablation time scales seem to be generally too long
for this process \citep{sbl+96}. Untill recently very few BW pulsars
were known in the Galactic field. The situation has changed in the
last few years thanks to deep radio follow-ups of gamma-ray point
sources detected by the LAT \citep{aaa+09} instrument on board the
{\it{Fermi}} satellite (see e.g. \citealt{rap+12} and references therein),
through which also binary pulsars whose radio signal was eclipsed for
a large fraction of the orbit were discovered. We note that, in
average black widows detected through this channel are more energetic,
show longer eclipses and have lighter companions than those previously
known, hence it is likely that the mass loss rate is higher and the
time for complete ablation of their companions shorter.

Other possible explanations of the formation of isolated MSPs invoke
different paths: one example is the accretion-induced collapse of a
white dwarf \citep{bg90}, another is white dwarf mergers
\citep{mic87}. Finding intrinsic differences between the two MSP
populations might point to a different origin and/or evolutionary path
for the isolated MSPs.

MSPs are more stable clocks than the non-recycled pulsars (PSRs).
When in binary systems, their stable rotation allows precise
determination of the Keplerian orbital parameters and, in some cases,
also the post-Keplerian parameters, making these pulsars laboratories
for testing relativistic gravity in the strong field regime
(e.g. \citealt{sttw02, ksm+06, wnt10}). If the rotational stability
and timing precision are high enough, MSPs can be used in timing
arrays for gravitational wave detection \citep{hd83}.

The High Time Resolution Universe survey (HTRU; \citealt{kjv+10}),
whose observations started at the Parkes observatory (NSW, Australia)
in 2008, was primarily designed to discover MSPs missed in previous
surveys. Thanks to the new digital backends adopted in this
experiment, the time and frequency resolutions have improved by a factor
of 4 and 8 respectively. This allows the survey to sample a much
larger volume of the Galaxy for short period pulsars as the
deleterious effects of dispersion are greatly diminished. The
suitability of the chosen parameters of the survey to discover MSPs
has been already demonstrated in \citet{bbb+11} where we presented the
first five new objects of this class found in the intermediate
latitude ($|b| < 15^\circ$) part of the survey. In this paper we
present the discovery and timing parameters of five further such
objects, confirming the very high potential of our acquisition system.

In section \S \ref{sec:time} we briefly describe the follow-up
observations and report the spin, orbital and astrometric
parameters resulting from the timing campaign on the five discovered
MSPs, highlighting their potential as `test particles' in the timing
array experiments. Given the increased sample of recycled pulsars, in
\S \ref{sec:lum} we revise the luminosity distribution of isolated
versus binary MSPs and find them to be very different. 
In \S \ref{sec:mw} we briefly report on the search
for gamma-ray counterparts to our discoveries. Finally, in \S
\ref{sec:pol} we describe the polarimetric characteristics for four of
the new discoveries.

\section{Observations and Timing Analysis} 
\label{sec:time}

Of the five newly discovered MSPs, four were found in the data of the
intermediate latitude part of the HTRU survey, consisting of 540s
pointings covering the area enclosed by Galactic coordinates
$-120^\circ < l < 30^\circ$, $|b| < 15^\circ$. The fifth,
J2236$-$5527, was discovered in the high latitude part of the survey:
270s pointings covering the entire southern sky below declination
$+10^{\circ}$ (except the intermediate latitude part mentioned above).

When a pulsar is discovered only its approximate spin period,
dispersion measure (DM, i.e. the free electron column density along
the line of sight) and position are known. To ensure that the full
potential of these discoveries is exploited, a precise determination
of the pulsars' position is needed (e.g. to look for counterparts at
other wavelengths; see \S \ref{sec:mw}) and accurate measurements of
the spin, astrometric, and if they belong to a binary system, orbital
parameters are necessary. To obtain these parameters one has to start
a follow-up timing campaign covering a time span of at least one year.

For the HTRU MSPs, confirmations, gridding observations (aimed at
determining a more precise position for the pulsar with a grid of 5
adjacent pointings; \citealt{mhl+02}) and the first timing
observations, are done with the same instrumentation used for the
survey pointings (see \citealt{kjv+10}). This allows the raw data to
be retained with all the time samples and the frequency channels, in
such a way as to be able to recover the periodic signal even if the
apparent spin period is significantly different from the discovery one
because of orbital motion. The rest of the timing observations, done
with an approximately monthly cadence, are performed either with the
Parkes 64-m telescope or, for the objects visible from the Northern
hemisphere, with the Lovell Telescope (Jodrell Bank, Manchester,
UK). At both sites an ATNF digital filterbank (DFB) is used as backend
and most of the observations are performed at a wavelength close to 20
cm. In particular, at the Parkes telescope the data are acquired over
256 MHz of bandwidth centered at a frequency of 1369 MHz and split
into 1024 frequency channels, while the Lovell data cover a frequency
range of 384 MHz centred at 1532 MHz and split into 1024 frequency
channels, more than 1/4 of which, untill September 2012, were removed
trough a bandpass filter because of Radio Frequency Interferences
(RFI). After that date the Lovell data have been manuallly cleaned
from RFI and a larger part of the band has been used. DFB data are
folded on-line using the pulsar ephemeris obtained from the first
stages of timing and refined as the follow-up observations
proceed. Observations of PSR\,J1832$-$0836 at Jodrell Bank were
performed also using the \textit{ROACH} pulsar backend. This backend
uses a ROACH FPGA
board\footnote{\ttfamily{https://casper.berkeley.edu/wiki/ROACH}} to
perform either baseband recording or coherent dedispersion of two
8-bit Nyquist sampled orthogonal polarisations over a total bandwidth
of 512\,MHz. After sampling and digitization, the ROACH board
channelizes the signal into 32 sub-bands of 16\,MHz, each of which is
coherently dedispersed and folded using an 8-core Intel Xeon computer
running the \texttt{dspsr} software package \citep{vb11}. As the total
usable bandwidth of the L-band receiver is 400\,MHz, the data from 25
subbands is combined and analyzed with the \texttt{psrchive}
package\footnote{\ttfamily{http://psrchive.sourceforge.net/}}
\citep{hvm04}. Also in this case about 1/4 of the bandwidth was
removed because of RFI untill September 2012. After about one year of
timing observations the covariance in the timing parameters is broken
and a precise timing solution is obtained. Times of arrival are
obtained using {\ttfamily{pat}} from {\ttfamily{psrchive}} through
cross-correlation with an analytic profile template created by fitting
von Mises functions to profile components.

Table \ref{tab1} reports the main detection characteristics of the
five newly discovered pulsars: for each object the timing position in
Galactic coordinates, the beam in which the pulsar was detected, the
radial distance of the pulsar from the discovery beam centre in units
of the beam radius, the signal-to-noise ratio (S/N) of the discovery
observation, the mean flux density $S$ and the pulse widths at 50\%
and 10\% of the peak of the mean pulse profile are listed. For the
three pulsars timed at the Parkes observatory, the values of the
fluxes are properly calibrated on a reference source (see \S
\ref{sec:pol}). For polarisation measurement purposes (see \S
\ref{sec:pol}) J1832$-$0836, timed over more than one year with the
Lovell telescope, was also observed on one occasion for $\sim 1.5$ hrs
with the Parkes Radio Telescope and hence as also properly flux
calibrated.  The flux density value for the remaining pulsar has been
obtained using the radiometer equation (see e.g. \citealt{mlc+01})
scaled using the calibrated flux densities of the other four MSPs. For
J1545$-$4550, for which we have a flux measurement also at 10 cm and
40 cm, we could calculate a spectral index $\alpha$ (with $S \propto
\nu^{-\alpha}$) of $1.15 \pm 0.09$ for the main peak, and a steeper,
though less constrained, spectral index of $2.0 \pm 0.6$ for the
second peak.

\begin{table*}
\caption{From left to right, we report, for our 5 MSPs, the name, the
  Galactic coordinates obtained from timing, offset with respect to
  the discovery position in beam radii, signal-to-noise ratio at discovery, mean flux
  density and pulse widths measured at 10\% and 50\% of the peak. For
  the pulsars with multiple profiles the pulse widths refer to the
  main peak alone. For J1545$-$4550 we report the flux density $S$ and
  pulse widths values at 10 cm$^a$ and 40 cm$^b$ as well. All the
  other numbers refer to the 20-cm observations. The numbers in
  parentheses are the 2-$\sigma$ errors on the last quoted digit(s). A
  30\% error on the flux density is assumed for the uncalibrated
  measurement.}
\label{tab1}
\begin{tabular}{lrrcccccc}
\hline
Name &  l$\;\;$$\;\;$   & b$\;\;$  & $\Delta_{\rm{pos}}$ & beam & S/N & $S$ & W$_{10}$ & W$_{50}$\\
 & ($\circ$)$\;\;$  & ($\circ$) & & (beam radii) & & (mJy) & (ms) & (ms) \\
\hline
J1431$-$5740  & 315.96 & 2.66 & 0.47 & 11  & 10.9 & 0.351(8) & 0.482 & 0.176 \\
J1545$-$4550  & 331.89 & 6.99 & 0.69 & 10  & 16.6 & 0.752(6) & 0.299 & 0.128 \\
J1545$-$4550$^a$ &     &      &  --  &  -- &   -- & 0.209(6) & 0.242 & 0.084 \\
J1545$-$4550$^b$ &     &      &  --  &  -- &   -- & 1.94(4)  & 0.746 & 0.284 \\
J1825$-$0319  & 27.05  & 4.14 & 0.38 & 12  & 11.1 & 0.20(6)  & 0.385 & 0.149 \\
J1832$-$0836  & 23.11  & 0.26 & 0.43 & 9   & 13.9 & 1.10(2)  & 0.339 & 0.058 \\
J2236$-$5527  & 334.17 & $-$52.72 & 0.67 & 1 & 27.8 & 0.282(8) & 0.979 & 0.195 \\
\hline
\end{tabular}
\end{table*}

\subsection{Binary pulsars}
\label{sec:bin}

\begin{figure}
\includegraphics[width=8truecm]{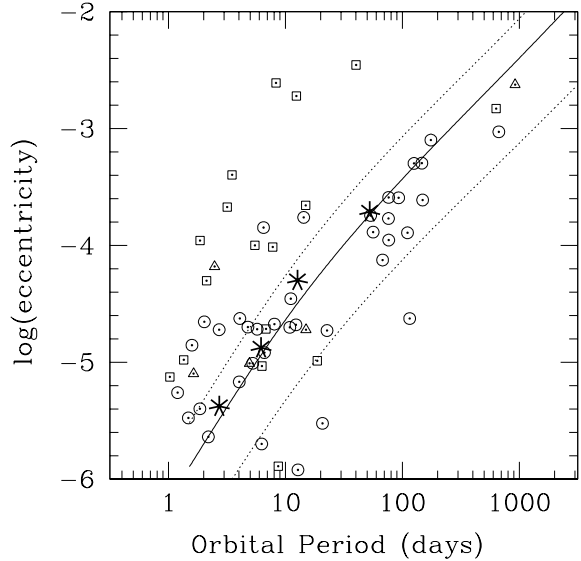}
\caption{Orbital eccentricity vs orbital period of binary pulsars in
  the disk of the Galaxy with measured eccentricities $e < 0.1$ (data
  taken from the ATNF pulsar catalogue; \citealt{mhth05}). The dotted
  lines contain 95\% of the eccentricities of He-WD MSPs according to
  the model of \citet{pk94}. The binary pulsars from this work are
  represented by black asterisks. Squares are Intermediate Mass Binary
  Pulsars and circles are Low Mass Binary Pulsars, while triangles
  denote pulsars not falling in these categories (e.g. the HTRU Very
  Low Mass Binary pulsar J1502$-$6752, \citealt{kjb+12}) or whose
  classification is still uncertain
  \citep{clm+01,eb01b,lfl+06,jsb+10,kjb+12}.}
 \label{fig:pbecc}
\end{figure}

Four of the five newly discovered pulsars orbit companions with
minimum masses ranging from 0.15 M$_{\odot}$ to 0.22 M$_{\odot}$ (see
Table \ref{tab2}). This mass range suggests that the companion stars
are Helium white dwarfs (He-WD), a hypothesis supported by the fact that
their measured orbital periods and eccentricities (Tab. \ref{tab2})
nicely follow the orbital relation predicted by \citet{phi92} for Low Mass
Binary Pulsars with He-WDs companions (see Figure \ref{fig:pbecc}).

The ephemerides resulting from the timing campaign of the binary MSPs
J1431$-$5740, J1545$-$4550, J1825$-$0319 and J2236$-$5527 are obtained
with {\ttfamily{tempo2}} \citep{hem06} and are presented in Table
\ref{tab2}. In the cases of J1545$-$4550 and J2236$-$5527,
polarisation calibration (see \S \ref{sec:pol}) allowed to further
improve the rms of the residuals of the timing solution with respect
to that obtained using uncalibrated data, by 5\% and 12\%
respectively. We note that, here and in Table \ref{tab:iso}, the
values for $\dot{P}$ (and the parameters derived from it) are not
corrected for the contributions of the acceleration along the line of
sight due to the potential well of the Galaxy and of the unknown
transverse velocity of the pulsar \citep{shk70}. Even though our
current data span is not long enough to measure a proper motion, if we
assume an average velocity of 70 km/s \citep{hllk05}, we obtain
contributions to the measured $\dot{P}$s of 2 to 18 per cent,
depending on the source. The uncertainty in the distances to the
pulsars also contributes to the uncertainty in the intrinsic value of
the spin-down. The listed values of the spin period derivative and of
the derived quantities should hence be taken as upper limits (lower
limits, in the case of the age).

\begin{table*}
\begin{center}
\caption{Timing parameters for the four binary pulsars discovered. In
  the top section of the table the measured astrometric, rotational
  and orbital parameters are listed: J2000 right ascension and
  declination obtained with the DE405 solar system ephemeris, spin
  period, period derivative, reference epoch for the spin period,
  dispersion measure, orbital period, projected semi-major axis, epoch
  of periastron, orbital eccentricity and longitude of periastron. The
  ELL1 binary model \citep{wex98} was used to fit such low
  eccentricities and the values reported have been derived from the
  EPS1, EPS2 and TASC parameters given by the model. The second part
  reports the derived parameters: minimum companion mass, implied
  dipolar magnetic field strength, characteristic age, spin-down
  energy, distance as derived from the DM and the NE2001 model
  \citep{cl02} for the distribution of free electrons in the Galaxy,
  and the luminosity at 20 cm (calculated as $S \times Dist^2$). The
  last part lists the time span covered by the timing data, the final
  timing solution rms residuals, the factor EFAC for which the times
  of arrival were multiplied to obtain a reduced $\chi^2$ of 1 and the
  number of times of arrival used. Numbers in parentheses are twice
  the formal {\ttfamily{tempo2}} errors on the last quoted
  digit(s). To obtain these ephemerides, as well as those in table
  \ref{tab:iso} the Barycentric Coordinate Time TCB, the standard for
  {\ttfamily{tempo2}}, was used.}
\label{tab2}
\begin{tabular}{lllll}
\hline
                     & J1431$-$5740       & J1545$-$4550         & J1825$-$0319      & J2236$-$5527       \\
\hline
Ra  (h m s) & 14:31:03.4953(3)          & 15:45:55.94596(4) & 18:25:55.9531(5) & 22:36:51.8510(8) \\
Dec ($\circ$ $\prime$ $\prime\prime$)   & $-$57:40:11.670(4)   & $-$45:50:37.5272(8)    & $-$03:19:57.570(15) & $-$55:27:48.833(4)   \\
P (ms)              & 4.1105439567658(12)& 3.57528861884712(12) & 4.553527919736(4) & 6.907549392921(3)  \\
$\dot{\rm{P}}$ (s/s) & 6.42(12)$\;\times 10^{-21}$       & 5.250(3)$\;\times 10^{-20}$         & 6.8(3)$\;\times 10^{-21}$         & 9.6(6)$\;\times 10^{-21}$          \\
Pepoch (MJD)         & 55865              & 55937                & 55784             & 55758              \\
DM (pc$\;$cm$^{-3}$)       & 131.46(3)          & 68.390(8)            & 119.5(4)          & 20.00(5)           \\
$P_b$ (days)         & 2.726855823(16)    & 6.203064928(8)       & 52.6304992(16)    & 12.68918715(14)    \\
a$sin$i (lt-s)       & 2.269890(4)        & 3.8469053(6)         & 18.266400(11)     & 8.775877(7)        \\
T$_0$ (MJD)          & 55461.79(27)       & 55611.40(2)          & 55805.06(5)       & 55472.551(7)       \\
e                    & 4.3(28)$\;\times 10^{-6}$          & 1.30(4)$\;\times 10^{-5}$            & 1.939(12)$\;\times 10^{-4}$       & 5.02(18)$\;\times 10^{-5}$         \\
$\omega$ (deg)       & 95(40)             & 221.3(18)            & 93.2(4)           & 350.8(18)          \\
\hline
M$_{2,\rm{min}}$ (M$_\odot$) & 0.156         & 0.153                & 0.177             & 0.223              \\ 
B (G)                & 1.64$\;\times 10^8$           & 4.38$\;\times 10^8$             & 1.78$\;\times 10^8$          & 2.61$\;\times 10^8$           \\
$\tau_c$ (yr)        & 1.01$\;\times 10^{10}$           & 1.08$\;\times 10^9$             & 1.06$\;\times 10^{10}$          & 1.14$\;\times 10^{10}$           \\
$\dot{\rm{E}}$ (erg$\;$s$^{-1}$) & 3.6$\;\times 10^{33}$          & 4.5$\;\times 10^{34}$              & 2.9$\;\times 10^{33}$           & 1.2$\;\times 10^{33}$            \\
Dist (kpc)           & 2.6                & 2.1                  & 3.1               & 0.8                \\
L$_{1400}$ (mJy$\;$kpc$^2$) & 2.4         & 3.3                  & 1.9               & 0.2                \\
\hline
Start (MJD)          & 55539.861          & 55685.428            & 55267.462         & 55504.474          \\
Finish (MJD)         & 56190.182          & 56190.238            & 56159.946         & 56013.031          \\
Data span (yr)       & 1.8                & 1.4                  & 2.4               & 1.4                \\
rms ($\mu$s)         & 5.638              & 0.825                & 21.273            & 5.355              \\
EFAC                 & 1.021              & 0.7828               & 1.291             & 0.985              \\
NToA                 & 45                 & 30                   & 61                & 35                 \\
\hline
\end{tabular}
\end{center}
\end{table*}

The timing residuals for J1545$-$4550 have an rms smaller than 1
$\mu$s, demonstrating why this pulsar has, as of March 2012, been
included in the Parkes Pulsar Timing Array (PPTA;
e.g. \citealt{mhb+12}) for the detection of stochastic gravitational
wave background. PPTA observations are performed at intervals of 2 - 3
weeks in three frequency bands: besides the 20-cm dataset collected
starting in May 2011, for J1545$-$4550 we hence also have a small
number of 10-cm and 40-cm observations (central frequencies of 3100
and 732 MHz respectively and bandwidths of 1024 an 64 MHz
respectively, split into 1024 channels). Profiles at the three
frequencies are shown in Fig. \ref{fig:J1546}.

\subsection{The isolated MSP J1832$-$0836}

Timing results from 1.3 years of follow-up observations of the
isolated MSP J1832$-$0836 with the Lovell telescope are presented in
Table \ref{tab:iso}. As for the binary MSPs, the $\dot{P}$ value is
not corrected for the contributions of the acceleration along the line
of sight due to the potential well of the Galaxy and of the transverse
velocity of the pulsar.

\begin{table}
\begin{center}
\caption{Ephemeris for the isolated pulsar J1832$-$0836. The
parameters are the same as in
Tab. \ref{tab2}.}
\label{tab:iso}
\begin{tabular}{ll}
\hline
                     &  J1832$-$0836      \\
\hline
Ra  (h m s)          & 18:32:27.59442(4) \\
Dec ($\circ$ $\prime$ $\prime\prime$)   & $-$08:36:54.54.969(3)   \\
P (ms)               & 2.7191120623197(3) \\
$\dot{\rm{P}}$ (s/s) & 8.62(5)$\;\times 10^{-21}$        \\
Pepoch (MJD)         & 55936              \\
DM (pc$\;$cm$^{-3}$)       & 28.18(9)           \\
\hline
B (G)                & 1.55$\;\times 10^8$           \\
$\tau_c$ (yr)        & 5.00e$\;\times 10^9$           \\
$\dot{\rm{E}}$ (erg$\;$s$^{-1}$) & 1.7$\;\times 10^{34}$          \\
Dist (kpc)           & 1.1                \\
L$_{1400}$ (mJy$\;$kpc$^2$) & 1.3         \\
\hline
Start (MJD)          & 55691.211          \\
Finish (MJD)         & 56188.933          \\
Data span (yr)       & 1.3                \\
rms ($\mu$s)         & 2.0                \\
EFAC                 & 1.382              \\
NToA                 & 43                 \\
\hline
\end{tabular}
\end{center}
\end{table}

The residuals of this pulsar have the third-smallest rms (2.0 $\mu$s;
see Tab. \ref{tab:iso}) among the HTRU MSPs discovered so far. This
target is hence another good candidate for Timing Array studies and
may be included in the target list for the European and North American
Pulsar Timing Arrays (EPTA; \citealt{fvb+10} and NANOGrav;
\citealt{jfl+09}), also considering that the current results are
likely to be improved, having been obtained with relatively short
timing observations (20 minutes or less over a clean bandwidth of
$\lsim 300$ MHz) and with no polarisation calibration.

This pulsar, with its 1400-MHz luminosity of 1.3 mJy$\;$kpc$^2$ (2.2
mJy$\;$kpc$^2$ if we use the distance obtained with the \citealt{tc93}
electron density model), is among the brightest isolated MSPs yet
discovered.

\section{MSP luminosity distribution}
\label{sec:lum}

Prompted by the case of J1832$-$0836, we briefly revise here the
comparison between the luminosity of isolated and binary MSPs, taking
advantage of the fact that in the recent years the number of Galactic
field MSPs, both isolated and in binary systems, has increased
significantly.

\citet{lmcs07} showed that the apparent differences in the luminosity
distributions seen in isolated and binary MSPs selected from the early
430-MHz surveys \citep{bjb+97} could be explained by small-number
statistics and observational selection biases. An examination of a
sample of 33 objects (11 isolated and 22 binaries) from 1400-MHz
surveys showed no statistically significant differences in the
distributions: the significance level of a Kolmogorov-Smirnov test (KS
test; see e.g. \citealt{pftv86}) was a mere 70.9\% (while it was
99.1\% for the 430-MHz sample). \citet{lmcs07} concluded that there
was no intrinsic difference in the two MSPs populations.

\begin{figure}
\includegraphics[bb=42 174 280 690, clip, width=8cm]{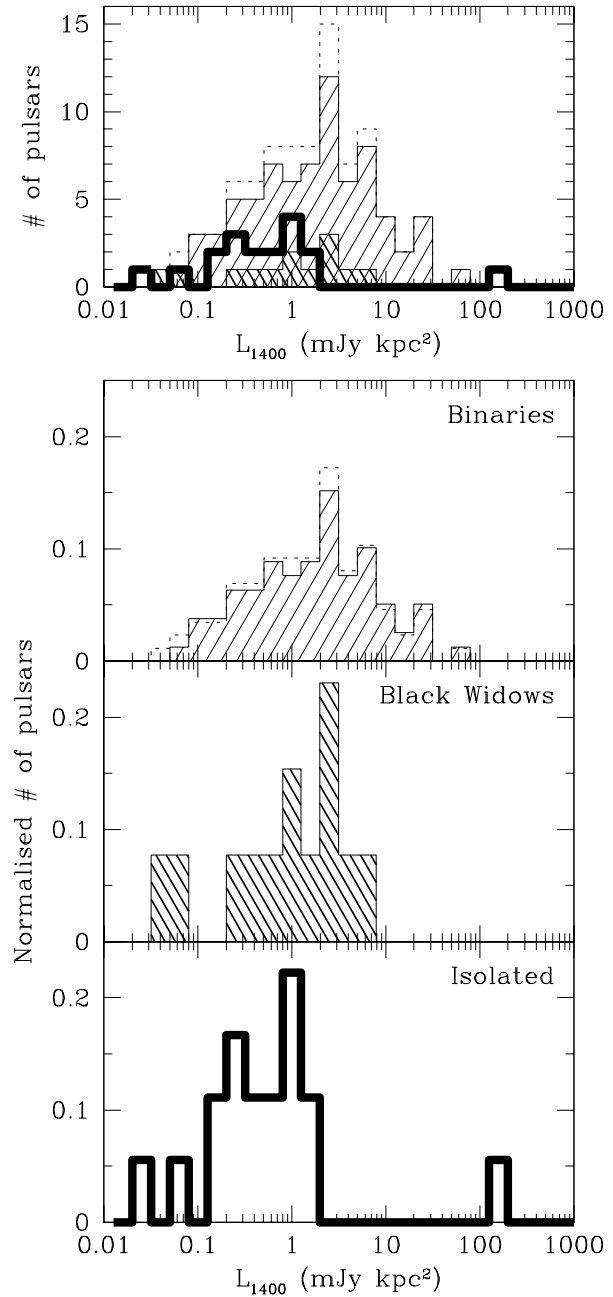}
\caption{1400-MHz luminosity distributions for a sample of 105
Galactic recycled pulsars. The dotted histogram shows the distribution
for 87 binary MSPs, 13 of which, included in the more densely shaded
histogram, belong to the class of black widow pulsars. The less
densely shaded histogram contains the other binaries. The thick line
shows the distribution for 18 isolated MSPs. In the top panel all the
MSP populations are shown unscaled. In the bottom three panels
binaries, black widows and isolated distributions have been normalised
to the total number of pulsars of each type. Flux densities at
1400-MHz are taken from the ATNF pulsar catalogue \citep{mhth05},
Thornton et al (in preparation)
and this paper. For six black widows flux densities (scaled at
1400-MHz with a spectral index $\alpha=1.8$) are taken from
\citet{hrm+11}, \citet{rrc+11}, \citet{rbg12} and \citet{rrc+13}. The
luminosities are computed using the distance from the NE2001 model
\citep{cl02}.}\label{fig:lum}
\end{figure}

Figure \ref{fig:lum} shows the updated distributions of the 1400 MHz
luminosity for 87 Galactic binary MSPs (dotted histogram; 13 in the
smaller shaded histogram are black widow pulsars, while the remaining
binaries are contained in the bigger shaded region) and 18 isolated
MSPs (thick line histogram). We here define MSP as all the recycled
pulsars having spin period $P < 0.1$ s and surface magnetic field $B <
2\times10^{10}$ G.  The results reported below do not significantly change even
if we consider only the fully recycled MSPs with $P < 0.01$ s. The
distances adopted for the calculation of the luminosities have been
obtained using the NE2001 model \citep{cl02} for the distribution of
free electrons in the Galaxy. A KS test on these two populations
suggests, with 99.9\% probability, that the two distributions
differ. Using the distances from the \citet{tc93} model the
probability is still 98.6\% and is again 99.9\% if we adopt the
modified version of the \citet{tc93} model, found to more closely
reproduce the distances of pulsars with measured parallaxes
\citep{sch12}. This result is in accordance with the earlier findings
by \citet{bjb+97}, \citet{kxl+98} and \citet{lkn+06}, but in contrast
with the aforementioned more recent work of \citet{lmcs07}, which had
a much smaller sample than ours. No difference in the period, period
derivative nor spin-down energy is seen between the two populations
(KS probabilities of 47.9\%, 47.9\% and 72.0\% respectively, and even
smaller considering fully recycled MSPs only).

One could argue that the discrepancy in luminosity can arise from the
fact that, in the current analysis of HTRU data (and in some of the
previuos surveys), no correction for the pulsar orbital motion through
acceleration searches (see e.g. \citealt{fsk+04,rem02,eklk13}) has
been done, leading to a loss of faint tight binaries in the sample
used. This could have enhanced the discrepancy in luminosity between
binary and isolated MSPs. However, if we restrict our analysis to
binary MSPs with orbital periods greater than 1 day, easily detectable
without any acceleration search in our 9 minutes observations, we end
up with a sample of 65 binaries whose luminosity distribution still
differs from that of isolated MSP with a probability of 99.9\%,
confirming our previous statement. We note also that the residual
observational bias that could be affecting the detection of tighter
system will be overcome in the near future: acceleration searches of
the HTRU data are in fact under way for the low-latitude section of
the survey (Ng et al. in preparation) and are about to be performed,
with the use of GPUs, on the entire mid-latitude section.

We therefore claim that the intrinsic luminosity functions are indeed
different, and may reflect differences in the evolutionary history of
isolated and binary MSPs. In the framework of the recycling scenario,
for instance, the progenitors of binary MSPs and isolated MSPs could
have experienced different kinds of accretion phases (this is
certainly true for the DRPs -- only 3 in our sample -- which are
subjected to a smaller amount of mass transfer), as also speculated by
\citet{kxl+98}. In particular, if isolated MSPs are the final outcome
of black widow pulsars that completely ablate their companions
(we hence exclude DRPs in this case), we could imagine that, during
the mass transfer in the ultracompact low-mass X-ray binary leading to
the BW formation, some modification is left in their magnetosphere
which is later reflected on their radio luminosity.

If this was the case we would expect the luminosity distribution of
BWs to be close to that of isolated MSPs and to differ from that of
pulsars in other binary systems. A KS test comparing the 1400-MHz
luminosity of our sample of binaries (excluding BWs) with that of 13
black widow pulsars (smaller, more densely shaded histogram in
Fig. \ref{fig:lum}) gives, however, a probability of only 59.8\% that
the two distributions derive from different populations (the
probability goes up to 87.7\% if we compare BWs to isolated MSPs). We
note however that, besides the small number statistics and the errors
introduced in the luminosity calculation by the poor knowledge of the
distances, in the case of the BW sample there is an additional
uncertainty affecting the results: about a half of the flux densities
reported in the literature \citep{hrm+11,rrc+11,rbg12,rrc+13}, in
fact, have been measured at a frequency other than 1400-MHz and have
been scaled to 1400-MHz using a theoretical spectral index
$\alpha=1.8$.

If, as the above considerations seem to suggest, the details of the
accretion phases are hardly responsible for the differences in
luminosity, another possibility, continuing with the assumption that
isolated MSPs derive from BWs, is that the time span necessary to
achieve the complete ablation of the companion is very long and
consequently the population of isolated MSPs should be entirely made
of old objects (on average older than the BWs), while the rest of the
binaries should cover a wider span of ages, from the younger more
luminous ones to the older ones with luminosities similar to those of
the isolated MSPs. A correlation between luminosity and age might
hence exist. We note however that such a correlation is extremely
difficult (if not impossible) to be properly tested, given that the
characteristic age estimate usually adopted for pulsars is not a
reliable indicator of the true age for MSPs (see
e.g. \citealt{tlk12}). For the binaries one could also try to use the
WD companions cooling age estimates \citep{hp98a}, but this indicator
has large uncertainties as well.

A further possibility, of course, is that the recycling scenario does
not apply to isolated MSPs whose different luminosity distribution
would be the consequence of a completely different formation
mechanism. If they were born directly as isoltaed objects, for
instance, we could expect that their post-Supernova velocities were in
average bigger than that of binary systems; consequently we'd expect
that their heights ZZ above the Galactic plane were bigger than for
binaries. A KS test on the ZZ distributions of our binary and isolated
MSP sample, however, does not support this hypotesis.

To further check the idea of different formation histories being the
cause of the different distributions, we also compared the luminosity
of isolated and binary MSPs in Globular Clusters. In these dense
environments, in fact, isolated MSPs are extremely likely to be
created by disruption of already formed MSP binary systems by the
close encounter with another star \citep{ps91}: therefore, isolated
and binary MSPs are expected to have had the same evolution, and thus
comparable luminosity distributions. A KS test performed on the
Cluster MSPs listed in \citet{blc11} results in a 88\% probability
that the two populations are indeed equal, corroborating the idea that
a difference in luminosity reflects a difference in evolution.

Given the small number of field MSPs and the uncertainties in the
distances, however, the hypotheses discussed above are still quite
speculative and the discovery of more isolated MSPs is desirable in
order to give statistical strength to any of these theses.

PSR~J1832$-$0836 and PSR~J1729$-$2117 (Thornton et al. in preparation)
are the only two isolated MSPs found so far in the HTRU survey and
they represent a mere 7\% of HTRU MSP discoveries. This small
percentage is very different from the $\sim 25$\% of isolated pulsars
previously known in the sample of Galactic field MSPs and it is also
significantly smaller than the $\sim 12$\% of isolated objects present
among the MSPs discovered through gamma-ray observations (directly or
with follow-up radio observations; \citealt{rap+12}) thanks to the LAT
instrument \citep{aaa+09} on board the {\it{Fermi}} satellite. The
discrepancy between the Galactic isolated MSPs percentage resulting
from blind radio surveys and from the targeted follow up of {\it{Fermi}}
point sources can be explained by the combination of various effects:
the discoveries of unidentified {\it{Fermi}} sources and the subsequent
detection of radio binary MSPs in targeted searches, are not (or less)
affected by orbital smearing of the signal (this is not true for blind
searches done directly in gamma-rays; see e.g. \citealt{pgl+12}) or by
eclipses; also, the fact that some of the gamma-ray flux detected
can, in some cases, arise from off-pulse emission produced by
intra-binary shocks could enhance the binary detection rate. Binaries,
in summary, are less difficult to find than in a blind search in the
radio band. The extremely small percentage of HTRU isolated MSPs, on
the other hand, could be explained by a luminosity bias. Because of
the higher time and frequency resolution of our survey, we are more
sensitive to higher-DM MSPs than previous surveys. This means that we
are sampling a deeper volume of the Galaxy, especially towards lower
latitudes. If, as the histograms of Figure \ref{fig:lum} suggest,
isolated MSPs are on average less luminous than MSPs in binary
systems, going deeper in the Galaxy means being progressively more
sensitive to more luminous populations of objects: far away, dim
isolated MSPs are lost, while brighter binary MSPs are discovered also
deep into the Galactic plane. If we take into account the discoveries
in the intermediate latitude part of our survey (where most of our
MSPs have been found so far) we indeed see that the isolated MSPs
J1729$-$2117 (Thornton et al. in preparation)
and J1832$-$0836 are the second and third closest MSPs of our
sample. If J1832$-$0836 were at the average distance of 2.5 kpc of our
sample of binaries, its flux density would have been barely above the
survey sensitivity limit \citep{kjv+10}, and below it at the average
distance as derived from \citealt{tc93} distance model. As for
J1729$-$2117, its detection in the survey has been possible only
because of scintillation, since its average flux density is below the
survey sensitivity. The paucity and the characteristics of the HTRU
isolated MSPs, hence, further support the hypothesis that this family
of recycled objects is intrinsically less luminous than MSPs in binary
systems.

\section{Search for gamma-ray counterparts}
\label{sec:mw}

Prompted by the fact that, as for the vast majority (all but two) of
gamma-ray detected pulsars, our five MSPs have $\dot{E} >
3\times10^{33}$ erg$\;$s$^{-1}$ and $\sqrt{\dot{E}}/d^2 \gsim
10^{-16}$ (erg$\;$s$^{-1}$)$^{-1/2}\;$kpc$^{-2}$ (except for
J1825$-$0319), we searched for their gamma-ray counterparts in the
{\it{Fermi-LAT}} data, making use of the precise positions and of the
rotational and orbital ephemerides obtained in this work.
 
No obvious counterpart was found in the LAT 2-year Point Source
Catalog \citep{naa+12}. We hence folded the gamma-ray data using our
timing solutions and the {\ttfamily{fermi}} plugin \citep{rkp+11} for
{\ttfamily{tempo2}}, in order to enhance the signal above the
background. Two different attempts have been made, one with no energy
cutoff and taking event class 2 photons from within 5 degrees of our
targets, a second one selecting only photons with energy higher than
300 MeV and using an extraction radius of 1 degree, accounting for the
smaller point spread function at higher energies. In both cases we
excluded events with zenith angles $> 100^\circ$ to reject atmospheric
gamma-rays from the Earth's limb and we applied the temporal cuts
recommended for the galactic point sources in the {\it{Fermi}} Science
Support Centre web page, through the {\ttfamily{gtmktime}} tool.

No pulsating counterparts were found folding the entire {\it{Fermi}} data set
nor folding only the photons in the time span over which the radio
timing solution was obtained. This latter attempt was made to account
for possible, though unlikely, timing noise or glitching events
occurring outside our data span, that would have changed the phase of
arrival of gamma ray photons on a longer data set. The lack of
detection could be, at least for some of the sources, ascribed to the
fact that the real $\dot{P}$, hence the real $\dot{E}$ of these pulsar
is smaller than the observed one (see \S \ref{sec:bin}). Moreover,
using the distance obtained with the \citet{tc93} model, brings the
$\sqrt{\dot{E}}/d^2$ of four of our MSPs (except J1545$-$4550) below
the limit of $10^{-16}$ (erg$\;$s$^{-1}$)$^{-1/2}\;$kpc$^{-2}$ below
which only two MSPs have a detected pulsed gamma-ray counterpart.

\section{Polarimetry of four Millisecond Pulsars} 
\label{sec:pol}

In the following we present the polarisation analysis carried out for
four of our five MSPs. No polarisation calibration was possible for
the timing observations that we performed with the Lovell telescope;
given its peculiar pulse profile and its relative brightness, however,
we decided to obtain calibrated polarisation information also for
J1832$-$0836, to try and understand the geometry of its radio beam. We
hence observed it with the Parkes radio telescope for $\sim 1.5$ hr on
one occasion. For J1825$-$0319, on the other hand, getting a good
polarisation profile would have required a much longer ($\sim
25\times$) integration time. For this reason, and considering that its
single, relatively narrow pulse profile (see Fig \ref{fig:J1825}) would
likely lead to a poorly constraining measurement, we decided to
discard it from the polarimetric analysis.

To compensate for the different gains of the two feeds, every timing
pointing was calibrated using observations of a square wave obtained
within $\sim 0.5 - 1$ hour of it. We also calibrated the observations
in flux density using observations of Hydra A. Finally we corrected
for the variable relative feed orientation across the various
parallactic angles.

In order to obtain reliable results from the polarisation analysis, it
is necessary to account for the Faraday rotation across our 256 MHz
band, hence measuring and correcting for the Rotation Measure
(RM). The RM was derived as follows: first we equally divided the
total observed bandwidth in four sub-bands for every observation. We then
created integrated profiles for the four channels separately, summing
all the observations, each properly weighted according to its SNR. For
each channel, we obtained an average Polarisation Angle (PA) as in
Noutsos et al. (2008):
\begin{equation}
PA_{\rm{ave}}= \frac{1}{2} \arctan \left(\frac{\sum_{i=n_{\rm{start}}}^{n_{\rm{end}}} U_i}{\sum_{i=n_{\rm{start}}}^{n_{\rm{end}}} Q_i}\right)
\end{equation}
where $n_{\rm{start}}$ and $n_{\rm{end}}$ are the initial and the final bin
numbers of the pulse, $U_i$ and $Q_i$ are the two Stokes parameters
for the $i-th$ bin. Since we noticed that in J1431$-$5736 averaging
over the entire pulse compromised the quality of the RM fit, we
decided to derive the average PAs across the individual
components of the summed linearly polarised profiles independently. We
then performed a least squares fit to compute the RM:
\begin{equation}
{\rm{PA}}_{\rm{ave}}(f)={\rm{PA}}_{\rm{ref}} + {\rm{RM}}c^2\times \left(\frac{1}{f^2} - \frac{1}{f_{\rm{ref}}^2}\right)
\end{equation}
where PA$_{\rm{ave}}(f)$ is the PA$_{\rm{ave}}$ at a certain frequency $f$,
PA$_{\rm{ref}}$ is the PA$_{\rm{ave}}$ at a reference frequency $f_{\rm{ref}}$, and
$c$ is the speed of light (e.g. \citealt{lk05}).

As a final step, we corrected the observations for the computed RM and
we summed them obtaining the frequency integrated polarisation
profiles described in the following subsections. The main polarisation
parameters for the four pulsars presented are summarised in Table
\ref{tab:pol}.
 
\begin{table*}
\caption{Rotation measure, linear polarisation, net and absolute
  circular polarisations (in flux density and in percentage of the
  total intensity) for the three pulsars timed at Parkes and
  J1832$-$0836. The top part of the table refers to measurements
  obtained at 20 cm, while the second and third, for J1545$-$4550
  only, at 10 cm and 40 cm respectively. The peaks are numbered left
  to right in the plots, except for the IV peak of J1545$-$4550, being
  the left-most at 40 cm. The peaks with two blended components for
  J1832$-$0836 (around phases $-$0.05 and 0.3 in Fig. \ref{fig:J1832}),
  are computed as a single peak.}
\label{tab:pol}
\begin{tabular}{lcccrrcc}
\hline
Name &  RM  & $L$ & $L/S$  & $V\;\;\;$ & $V/S$ & $|V|$ &$|V|/S$ \\
 & (rad/m$^2$)  & (mJy) & (\%) & (mJy) & (\%) & (mJy) & (\%)  \\
\hline
\bf{20-cm}  &  &  &  &  &  &  & \\
\hline
J1431$-$5740  & $-50 \pm 15\;\;$ & 0.068(3)  & 19.3(8) & $-$0.088(4) & $-$25(1)   & 0.092(2) & 26.3(8) \\
J1545$-$4550  & $-0.6 \pm 1.3\;$     & 0.299(4)  & 53.1(4) & $-$0.094(2) & $-$16.6(4) & 0.094(1) & 16.7(3) \\
      peakII  & $-6.6 \pm 7.8\;\;$   & 0.064(1)  & 100(6)  &    0.009(2) &    15(4)   & 0.011(1) & 17(2)   \\
     peakIII  & $ -6.0 \pm 1.1\;\;$  & 0.074(1)  & 58(2)   & $-$0.018(3) & $-$14(2)   & 0.018(1) & 14(1)   \\
J1832$-$0836  & $25 \pm 17$            & 0.122(9)  & 25(2)   &    0.04(1)  &     8(3)   & 0.046(7) & 9.7(16) \\
     peakII   & --                     & 0.048(5)  & 16(2)   &    0.00(1)  &     0(3)   & 0.000(4) & 0.0(15) \\
     peakIII  & --                     & 0.048(7)  & 18(3)   & $-$0.023(8) &  $-$8(3)   & 0.038(5) & 14(2)   \\
     peakIV   & --                     & 0.049(7)  & 67(12)  &    0.012(8) &   17(12)   & 0.027(5) & 38(8)   \\
J2236$-$5527  & $27.8  \pm 1.4\;\;$  & 0.062(2)  & 22.1(8) & $-$0.001(3) &     0(1)   & 0.029(2) & 10.5(7) \\
\hline
\bf{10-cm}  &  &  &  &  &  &  & \\
\hline
J1545$-$4550  & $10.5 \pm 5.7$         & 0.121(1)  & 58(1)   & $-$0.022(3) & $-$11(1) & 0.022(2) & 10.5(8) \\
     peakII   & --                       & 0.010(2)  & 38(7)   & $-$0.002(2) & $-$8(9)  & 0.002(1) & 8(5) \\
\hline
\bf{40-cm}  &  &  &  &  &  &  & \\
\hline
J1545$-$4550  & $\;\;3 \pm 15$     & 0.37(2)  & 35(2)   & $-$0.17(3) & $-$17(3) & 0.17(2) & 17(2) \\
     peakII   & $-9.7 \pm 8.9\;\;\;$   & 0.34(1)  & 92(5)   &    0.05(2) &    12(6) & 0.05(1) & 12(3) \\
     peakIII  & $-0.7 \pm 0.6\;\;\;$   & 0.15(1)  & 100(25) &  0.001(23) &    1(21) & 0.02(1) & 14(12) \\
     peakIV   & --                       & 0.14(2)  & 33(5)   &    0.02(3) &     6(8) & 0.05(2) & 13(5) \\
\hline
\end{tabular}
\end{table*}

Polarisation studies are fundamental to investigate the geometry of
pulsars as well as their magnetic fields and radio beam structures. In
the ideal case, the interpretation of the results of this kind of
analysis could be carried out using the Rotating Vector Model (RVM) by
\citet{rc69a}. This model assumes that the position angle of the
linear polarisation ($L = \sqrt{Q^2 + U^2}$) is tied to that of the
pulsar magnetic field lines. In this way, its prediction is to observe
a swing of the Polarisation Angle (PA) curve across the pulse (that
should produce a characteristic ``S-shape''), due to the sweeping of
the observer's line of sight through the radio beam.

%

The fit of the model should permit the inclination between the magnetic
and the rotational axes and the impact parameter between the line of
sight and the magnetic axis to be obtained.


This however is not always possible, for example because the
polarisation signal-to-noise ratio (SNR) is not high enough to allow a
good determination of the PA, or the pulse is so narrow that there is
not enough information to constrain the fit, or also because the
predicted PA swing is not displayed (for a thorough discussion of the
difficulties and limitation of the RVM see \citealt{ew01}). MSP PAs,
in particular, often display jumps or discontinuities in the PA
profile: this is the case of our MSPs, as described below, for which
hence a fit using the RVM has not been possible. 

\subsection{J1431$-$5740}

\begin{figure}
\includegraphics[width=8truecm, height=8truecm]{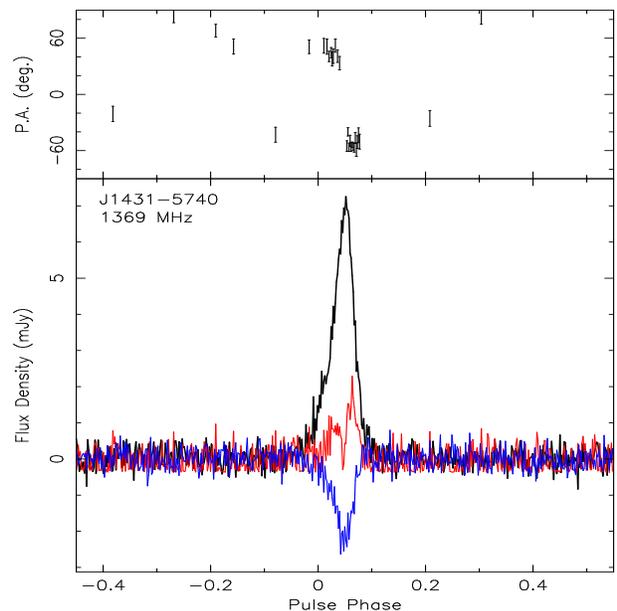}
\caption{Polarisation profile for J1431$-$5740 at a centre frequency
  of 1369 MHz. The lower panel shows, in black, the total intensity,
  in light grey (red in the electronic version) the linearly polarised
  intensity $L$ and in dark grey (blue in the electronic version) the
  circularly polarised intensity $V$. The top panel gives the PA of
  the linearly polarised emission.}
 \label{fig:J1431}
\end{figure}

\begin{figure}
\includegraphics[bb=298 55 705 496, clip, width=5.55truecm, height=8truecm, angle=270]{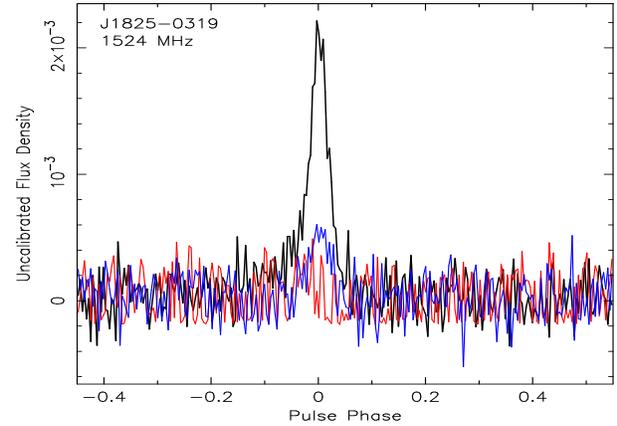}
\caption{Uncalibrated profile for J1825$-$0319 at a centre frequency
  of 1369 MHz. Lines and panels as in Figure \ref{fig:J1431}.}
 \label{fig:J1825}
\end{figure}

The pulse profile for J1431$-$5740 (Fig. \ref{fig:J1431}) shows a
single slightly asymmetric peak. The linear polarisation, however,
displays two peaks, whereof the leading is fainter and wider in
amplitude than the trailing. The minimum of the linear polarisation
occurs roughly at the phase of the peak in the total intensity
profile. The PAs in correspondence with the leading peak show a
slightly decreasing slope, while the PA curve of the trailing
component is flat. The two series of PAs exhibit an orthogonal (within
the uncertainties) jump in between. The circular polarisation profile
shows a single component, whose peak appears in phase with that of the
total intensity.

\subsection{J1545$-$4550}

\begin{figure}
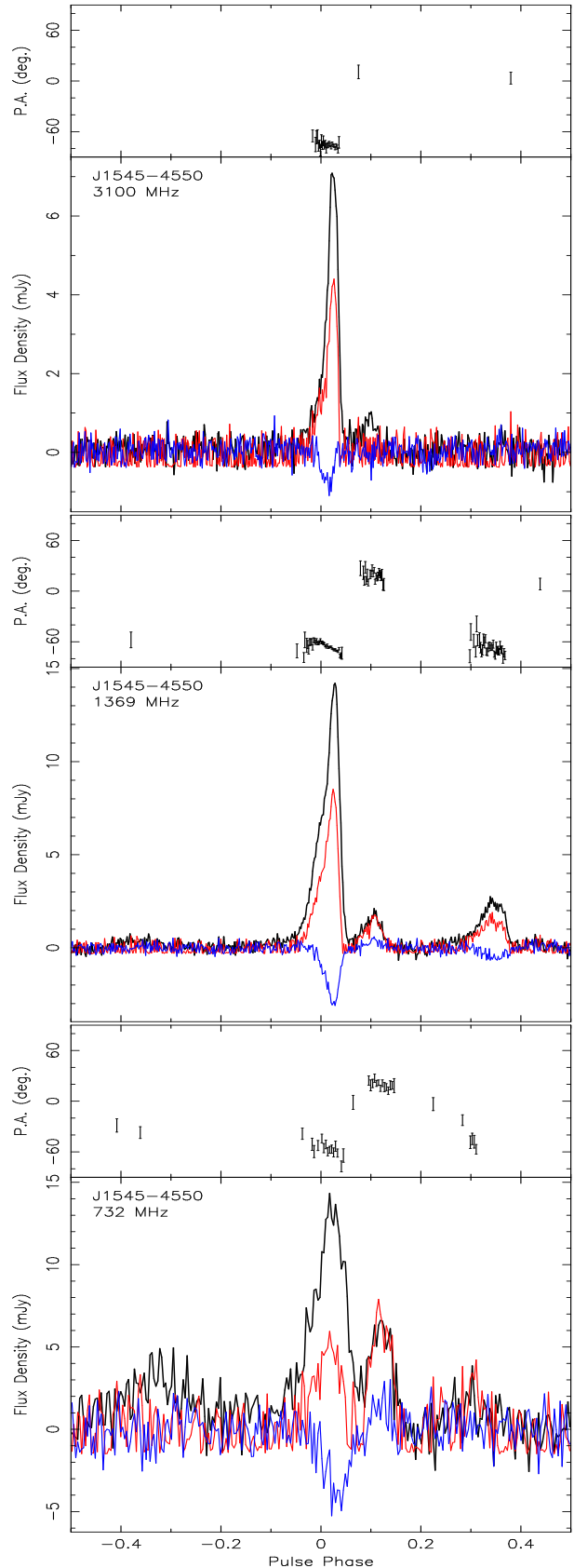

\includegraphics[bb=63 131 503 679, clip, width=8truecm, height=7.5truecm]{J1546-4552.10cm.3.ps}
\includegraphics[bb=63 131 503 661, clip, width=8truecm, height=7.25truecm]{J1546-4552.20cm.3.ps}
\includegraphics[bb=63 95 503 661, clip, width=8truecm, height=7.75truecm]{J1546-4552.50cm.3.ps}
\caption{Polarisation profiles for J1545$-$4550 at centre frequencies
  of 3100 MHz (top), 1369 MHz (centre) and 732 MHz (bottom). Lines and
  panels as in Figure \ref{fig:J1431}. The pulses at the different
  frequencies are aligned so that the centroid of the main peak
  falls at phase 0.}
 \label{fig:J1546}
\end{figure}

This pulse profile at 20 cm (central panel of Figure \ref{fig:J1546})
shows three well defined components: two of them are contiguous, while
the third one is 0.22 in rotational phase away from the closest
component. A fourth component is barely visible about 0.4 in phase
before the main peak. It acquires statistical significance when
comparing the 20-cm profile to the 40-cm one (bottom panel of Fig.
\ref{fig:J1546}) where this fourth component emerges clearly. The
linear polarisation profile follows the total intensity one, and the
percentage of $L$ under every component is very high, especially
beneath the second peak. The PA curve of the component with
higher SNR shows a short, steep rise and a smooth decline toward the
separation between the first and the second component. A similar
smooth decline with almost the same inclination is shown by the PA curve
of the third peak, while that of the second one is flatter
and exhibits a jump of $\sim90^\circ$ with respect to the first. The
circular polarisation appears mainly under the first component where
it is fully negative, while there are just hints of circular
polarisation beneath the other two peaks, with opposite signs.

As mentioned in \S \ref{sec:bin}, for this pulsar a few observations at
10 and 40 cm are also available. Given the smaller number of data
points, the resulting profiles, especially the one at 40 cm, whose
band is often corrupted by RFI, are not as high S/N as the 20-cm one.
At 10 cm (top panel of Figure \ref{fig:J1546}) the first peak appears
$\sim 35$ per cent narrower than at 20 cm and the third peak is
statistically undetectable. The linear and circular polarisation
profiles of the main peak behave as at 20 cm, while the second peak
appears to be unpolarised. The 40-cm profile (bottom panel of Figure
\ref{fig:J1546}) shows four components: besides the three clearly
present at 20 cm, whose peaks are closer together at this lower
frequency, a fourth component precedes in phase the main peak by
1/3. The percentage of linear polarisation of the main peak drops by
about 30 per cent with respect to the 20-cm value, while the second
and third peaks appear 100 per cent linearly polarised (even though
for the latter, as for the fourth peak, the statistic is quite
poor). The circular polarisation is negative below the main peak and
positive below the second one. In contrast with what is generally
found \citep{man71b, xkj+96}, for this object the percentage of linear
polarisation (at least in the main peak where a trend through all
three frequencies can be easily seen) decreases with decreasing
frequency while the percentage of circular polarisation increases (in
absolute value).

The pattern of the PA curves are very similar at the three frequencies
and the RM values calculated on the three data sets separately are
always compatible with zero at 2-$\sigma$.

\subsection{J1832$-$0836}

\begin{figure}
\includegraphics[width=8truecm, height=8truecm]{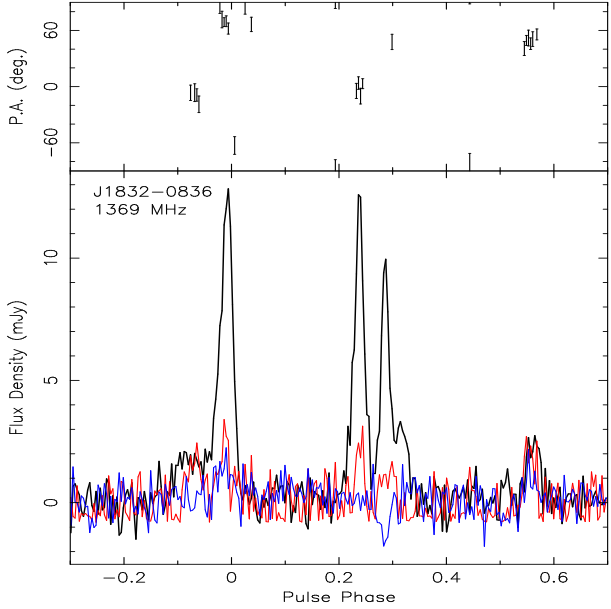}
\caption{Polarisation profiles for J1832$-$0836 at a centre frequency
  of 1369 MHz. Lines and panels as in Figure \ref{fig:J1431}.}
 \label{fig:J1832}
\end{figure}

This MSP shows a complex pulse profile displaying six components (see
Fig. \ref{fig:J1832}), two of which are blended with the adjacent
one. The polarisation values for the two peaks with the two blended
components are computed for both components together in Table
\ref{tab:pol}, listing hence a total of four peaks (centred
approximately around phases $-$0.05, 0.23, 0.3 and
0.57). Unfortunately the polarisation is quite small and the PAs do
not allow us to clearly understand the beam structure leading to such
a peculiar profile. The PAs in correspondence of the two components
around phase 0.0 are separated by a jump of almost $100^{\circ}$, with
the second component showing a slightly negative slope. In
contrast the peaks at phase $\sim0.25$ and $\sim0.57$ have PAs that
appear flatter, with a slightly positive slope for the latter.

The linear polarisation profile shows five peaks, with only the second
component of the peak around phase 0.3 apparently unpolarised. We
computed an RM value for every linearly polarised peak separately,
however, due to the poor signal-to-noise ratio of the polarised
emission, the errors are often very high and most of the RMs are
compatible with zero at 2-$\sigma$. The RM measured on the peak
centred at phase 0.0, the one showing the highest level of
polarisation and giving the smallest error is $\rm{RM}=35 \pm 2$
rad/m$^2$. To confirm this value we also measured the RM with the
methods described in \citet{njkk08} and in \citet{hml+06}, yielding
respectively values of $50 \pm 30$ rad/m$^2$ and $25 \pm 17$
rad/m$^2$. Testing these results with a matrix template matching
technique, we found out that a value close to 25 rad/m$^2$ better
describes the data.

The circular polarised emission is visible, with changing
signs, in correspondence with the peaks at phase $\sim0.0$, $\sim0.3$
and $\sim0.57$.

\subsection{J2236$-$5527}

\begin{figure}
\includegraphics[width=8truecm, height=8truecm]{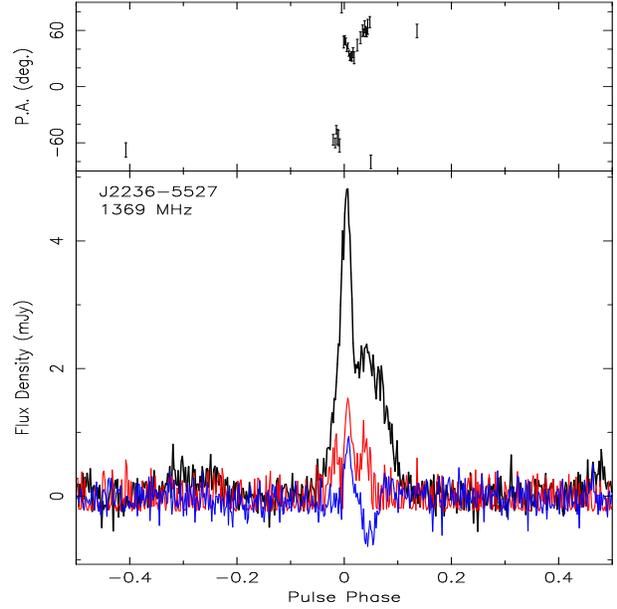}
\caption{Polarisation profiles for J2236$-$5527 at a centre frequency
  of 1369 MHz. Lines and panels as in Figure \ref{fig:J1431}.}
 \label{fig:J2236}
\end{figure}

The total intensity profile of this MSP (Fig. \ref{fig:J2236}) shows
two blended peaks and two additional much fainter components
separated in phase from the main one by --0.42 and 0.22
respectively. We considered just the main components, since there is
no evidence of polarised signal for the other two. The linear
polarisation profile shows three peaks. We computed the RM for each of
the peaks separately: the RMs of the first and the third peak seem to
be consistent with zero although with a very high uncertainty, while
the central peak gives a RM value of $\sim 27.8 \pm 1.3$
rad/m$^2$. The PAs of the first component draw a flat curve, and are
separated from the PAs related to the second peak by a jump of almost
$\sim100^\circ$. For the central component the PA curve shows a steep
decrease, whereas it increases in correspondence with the third
component. Circularly polarised emission is present mainly at the same
pulse longitude of the second and the third component of the linear
polarisation, with a $V$ profile sign change occurring at the same
time of the PA's slope change.

\section{Conclusions} 
\label{sec:conc}

In this paper we presented the discovery and timing results of five
MSPs detected in the HTRU survey data. Four of them belong to binary
systems with He-WD companions, while the fifth is one of the only two
isolated MSPs found so far in this survey. A revision of the
luminosity distribution of known isolated and binary MSPs is
presented; with the sample of isolated MSPs increased by more than 60
per cent and the total sample of MSPs tripled with respect to the
previous works addressing this issue (e.g. \citealt{lmcs07}), the
intrinsic luminosity difference between these two populations appears
to be confirmed. The small percentage of isolated objects among MSP
discoveries in the HTRU survey, sampling a deeper volume of the Galaxy
with respect to previous experiments, also supports the hypothesis
that the intrinsic luminosity difference is indeed real. This effect
could be the consequence of a different evolutionary history
experienced by isolated and binary MSPs.

For four of our targets, those for which we have follow-up observation
carried out at the Parkes radio telescope, we also reported the
results of a polarimetric analysis. The possibility to calibrate their
polarisation accurately also allowed us to obtain an improved timing
solution for two of our MSPs.

Thanks to the precise ephemerides obtained through radio timing, we
could fold in phase the gamma-ray photons collected by the {\it{Fermi-LAT}}
instrument since its launch from the direction of the
MSPs. Unfortunately no pulsating high-energy counterpart to any of our
MSPs was found, nor were point sources corresponding to the position
of our targets present in the 2-yr {\it{Fermi}} Point Source Catalog
\citep{naa+12}.

One of the main aims of the High Time Resolution Universe Pulsar
Survey \citep{kjv+10} is that of discovering stable millisecond
pulsars suitable to be included in pulsar timing array experiments for
the detection of gravitational waves \citep{hd83}. Two of the newly
discovered MSPs show quite small timing residuals, which makes them
good candidates to be used in pulsar timing arrays: J1545$-$4550 is
indeed the second HTRU MSP to be included in the Parkes PTA and
J1832$-$0836 is being considered for the European and North American
PTAs.

\section{Acknowledgements}
The Parkes radio telescope is part of the Australia Telescope which 
is funded by the Commonwealth of Australia for operation as a 
National Facility managed by CSIRO.

Part of this research was carried out at the Jet Propulsion
Laboratory, California Institute of Technology, under a contract with
the National Aeronautics and Space Administration.

MBu and AP have been partly funded under the ASI contract
I/047/08/0-WP3000 (Fermi).


\bsp
\label{lastpage}

\end{document}